\documentclass[prb,twocolumn,amsmath,amssymb,showpacs]{revtex4}
\usepackage{graphicx}

\def\beq{\begin{equation}}
\def\eeq{\end{equation}}
\def\beqa{\begin{eqnarray}}
\def\eeqa{\end{eqnarray}}

\begin{document}


\title{
Large-$N$ expansion based on the Hubbard operator 
path integral representation
and its application to the t-J model II. The case for finite $J$.
} 

\author{
Adriana Foussats and  Andr\'es Greco 
}

\affiliation{
Facultad de Ciencias Exactas, Ingenier\'{\i}a y Agrimensura 
and Instituto de F\'{\i}sica Rosario
(UNR-CONICET). 
Av. Pellegrini 250-2000 Rosario-Argentina.
}

\date{\today}

\begin{abstract}
We have introduced a new perturbative approach for $t-J-V$ model 
where Hubbard operators are treated as 
fundamental objects. 
Using our vertices and propagators we have developed 
a controllable large-$N$ expansion 
to calculate different correlation functions. We have investigated 
charge density-density response and the phase diagram of 
the model. 
The charge correlations functions are not very sensitive to the value
of $J$ and they show collective peaks (or zero sound) which are more 
pronounced when they are well separated (in energy) from the 
particle-hole continuum. For a given $J$ a Fermi liquid state is 
found to be stable for doping 
$\delta$  larger than a critical doping $\delta_c$. $\delta_c$ decreases 
with decreasing $J$. For the physical region of the parameters 
and, for $\delta< \delta_c$, the system enters 
in an incommensurate flux or 
DDW phase.
The inclusion of the nearest-neighbors Coulomb repulsion $V$ leads to a 
CDW phase when $V$ is larger than a critical value $V_c$. 
The dependence of $V_c$ 
with $\delta$ and $J$ is shown.
We have compared the results with other ones  in the literature.
\end{abstract}


\maketitle
\section{Introduction}

In the last years a large part of the solid state community has been devoted 
to understand the physics of strongly electronic correlated models.
The importance of this study is supported by  
clear experimental facts.
High-$T_c$ superconductors \cite{Anderson} and organic superconductors 
\cite{McKenzie} are considered, for instance, electronic correlated 
systems.
They show that  
electronic properties are very different from those expected in usual metals.

For better understanding  of the physics of these systems, 
it is very important
to develop different methods for studying models for correlated electrons.
The two dimensional ($2D$) $t-J$ model is probably one of the most simple 
models we should understand in the first place.  

The $t-J$ model is the strong coupling version of the Hubbard model
\cite{Izyumov}. In spite of its simple form, there is no exact solution 
for this model   
until now and,  
many analytical and numerical \cite{Dagotto} methods 
have been developed and the obtained results confronted to each other.

In the $t-J$ model double occupancy is forbidden and, 
the model can be written 
in terms of the correlated or projected operators,  also known  as 
Hubbard X-operators \cite{Hubbard}. 
Hubbard operators verify complicated commutation rules 
which are very different from the familiar commutation rules for usual 
fermions and bosons. 
One of the common methods, used for avoiding this problem, introduces 
slave-particles\cite{Izyumov} in order to decouple the original $X$-operator.
These slave-particles verify usual
commutation rules but, they are fictitious and 
sometimes it is not clear if the obtained results are
genuine or artifacts from merely decoupling. On the other hand,
the decoupling scheme introduces 
a gauge degree of freedom which requires a gauge fixing. 
The gauge fixing is a long discussed fundamental problem
(see for example Ref.[\onlinecite{Arrigoni}]). 

Another problem in the treatment of the $t-J$ model is 
the absence of any  small 
coupling parameter suitable for a perturbative expansion. 
To deal with this problem, strong-coupling techniques have been developed. 
Between them,
what will be important for the present paper is the large-$N$
expansion where $N$ is the number of electronic degrees of freedom per
site (see below). 
The large-$N$ expansion has been used extensively in the context of 
slave-boson approach \cite{Wang} 
(SBA) and, more recently, using Bayn-Kadanoff    
functional theory (BKF) in terms of $X$-operators \cite{Zeyher,Zeyher1}. 
 
On the basis of our Feynman path integral representation for the $t-J$
model \cite{Foussats1}, we developed a large-$N$ approach \cite{Foussats}
for the $J=0$ case 
($U$-infinite Hubbard model). 
This method has the advantage of working  with Hubbard operators as 
fundamental objects without any decoupling procedure. 
Then, we neither take care of   
gauge fluctuations nor Bose condensation as in the SBA.

The method developed in Ref.[\onlinecite{Foussats}] 
was recently applied   
to describe electronic 
properties of quarter-filled organic molecular crystals \cite{Merino}.
For instance, in Ref.[\onlinecite{Merino}] we have 
calculated electronic self-energies $\Sigma(k,\omega)$, spectral 
functions $A(k,\omega)$ and the electronic density of states $N(\omega)$
which are nontrivial calculations considering that they involve 
fluctuations, in a controllable approach, above the mean field. 
Our $A(k,\omega)$ and $N(\omega)$ were carefully 
compared \cite{Merino} with similar results 
obtained by Lanczos diagonalization. Good agreement has been found for 
the behavior of the dynamical properties at $J=0.0$.  

Our previous studies  for $J=0.0$ motivate us to 
extend our  approach to the case of finite $J$.  
On the other hand, many physical problems play with a finite 
value of $J$ and therefore,   
it is necessary to extend our approach to this case. 
This extension is the main topic of the present paper. 

As our method is new and we claim it can be used to simplify several
calculations, we must show: a) how our method explicitly works 
and, b) comparison of the results with others in the literature 
in order to show the confidence of the method. To satisfy a) and b) 
is one of the main purpose of the present paper.

In Sec.II we develop the perturbative expansion and we 
give the new Feynman rules which are used in the explicit 
calculation in Secs. III and IV. 
In Sec.III we show results for the charge-charge correlation functions. 
In Sec.IV we study different instabilities of the model. 
In Sec. V we give the conclusions.

\section{Perturbative approximation: A large-$N$ approach}
\label{stripfil}

In this section, we present the large-$N$ expansion for the $t-J-V$ model 
in the
framework of the path integral representation for Hubbard
operators. We will extend our formalism of Ref.[\onlinecite{Foussats}] to the case of 
finite $J$. 

As in Ref.[\onlinecite{Foussats}], our starting point for 
developing the large-$N$ approach is the  
path integral partition function written in the
Euclidean form  ($it \rightarrow \tau$),  

\begin{eqnarray}
Z&=&\int {\cal D}X_{i}^{\alpha \beta}\; \delta[X_{i}^{0 0} +
\sum_{\sigma} X_{i}^{\sigma \sigma}-1]\; \delta[X_{i}^{\sigma
\sigma'} - \frac{X_{i}^{\sigma 0} X_{i}^{0 \sigma'}}{X_{i}^{0 0}}]
\nonumber \\ &\times&(sdet M_{AB})_{i}^{\frac{1}{2}} exp\;(- \int
d\tau\;L_E(X, \dot{X}))\;.
\end{eqnarray}

In Eq.(1), 
the five Hubbard $X$-operators $\hat{X}^{\sigma
\sigma'}$ and $\hat{X}^{0 0}$ are boson-like and the four Hubbard ${\hat
X}$-operators $\hat{X}^{\sigma 0}$ and $\hat{X}^{0 \sigma}$ 
are fermion-like
\cite{Hubbard}. 
The spin index $\sigma$ is $\sigma = \pm$ ($up$ and $down$ state,
respectively).

There is also a superdeterminant  
$(sdet M_{AB})^{\frac{1}{2}} = 1/\frac{1}{(-X_{i}^{0 0})^{2}}$
formed with the set of all the second class constrains of
the theory 
(see Refs.[\onlinecite{Foussats}] and
[\onlinecite{Foussats1}]).

The Euclidean Lagrangian $L_E(X,\dot{X})$ in (1) is:

\begin{eqnarray}
L_E(X, \dot{X}) =  \frac{1}{2} \sum_{i,
\sigma}\frac{({\dot{X_{i}}}^{0 \sigma}\;X_{i}^{\sigma 0} +
{\dot{X_{i}}}^{\sigma 0}\; X_{i}^{0 \sigma})}
 {X_{i}^{0 0}}
+ H(X)\;.
\end{eqnarray}

On the basis of Hubbard $X$-operators, 
the $t-J-V$ Hamiltonian is of the form:

\begin{eqnarray}
H(X) &=& \sum_{<ij>,\sigma}\;(t_{ij}\; \hat{X}_{i}^{\sigma 0} 
\hat{X}_{j}^{0
\sigma} + h.c.) \nonumber \\ 
&+& 
\frac{1}{2} \sum_{<ij>;\sigma} J_{ij}
(\hat{X}_{i}^{\sigma
{\bar{\sigma}}} \hat{X}_{j}^{{\bar{\sigma}} \sigma} - 
\hat{X}_{i}^{\sigma
\sigma} \hat{X}_{j}^{\bar{\sigma} \bar{\sigma}})
 \nonumber\\
&+& \sum_{<ij>;\sigma \sigma'} V_{ij} \hat{X}_{i}^{\sigma \sigma}
\hat{X}_{j}^{\sigma' \sigma'}
-\mu\sum_{i,\sigma}\;\hat{X}_{i}^{\sigma \sigma}.
\end{eqnarray}

Now,  similarly to Ref.[\onlinecite{Foussats}]
we make the following changes in the path integral (1): 

a) We
integrate over the boson variables ${X^{\sigma \sigma'}}$ using
the second ${\delta}$ -function in (1).

b) The spin index $\sigma=\pm$, is extended to a new index $p$
running from $1$ to $N$. In order to get a finite theory in the
$N$-infinite limit,  we re-scale the hopping $t_{ij}$, 
the exchange parameters
 $J_{ij}$ and the nearest-neighbors Coulomb repulsion $V_{ij}$ to 
$t_{ij}/N$, $J_{ij}/N$ and $V_{ij}/N$, respectively.

c) The completeness condition ($X_{i}^{0 0} + \sum_p
X_{i}^{pp}=N/2$) can be exponentiated, as usual, by using the
Lagrangian multipliers $\lambda_i$.

d) The charge-like terms (the 3rd and 4th term of (3)) of the Hamiltonian,
after extended to large-$N$, are written in terms of $X^{00}$ using 
the completeness condition.    

e) 
We write $X^{00}$ and $\lambda$ in terms of static mean-field
values and dynamic fluctuations;
$X_{i}^{0 0} = N r_{0}(1 + \delta R_{i})$, 
$\lambda_{i} = \lambda_{0} + \delta{\lambda_{i}}$.

f) Finally, we make the following change of variables;
$f^{+}_{i p} = \frac{1}{\sqrt{N r_o}} X_{i}^{p 0}$,
$f_{i p} = \frac{1}{\sqrt{N r_o}} X_{i}^{0 p}$.

By following the steps a-f, we find the effective Lagrangian

\begin{widetext}
\begin{eqnarray}
L_{eff}& = & - \frac{1}{2}\sum_{i,p}\left(\dot{f_{i p}}f^{+}_{i p}
+ \dot{f^{+}_{i p}}f_{i p}\right) \frac{1}{(1 + \delta R_{i})} +
\sum_{<ij>,p}(t_{ij}\ r_{o}f^{+}_{i p}f_{j p} + h.c.)
 -   (\mu - \lambda_{0})\;\sum_{i,p}\;f^{+}_{i p}f_{i p}
\frac{1}{(1 + \delta R_{i})} \nonumber \\
&+&
N\;r_{0}\;\sum_{i}\delta{\lambda_{i}}\;\delta R_{i} 
+  \frac{1}{2}\sum_{<ij>,p,p'}\frac{J_{ij}}{N}\frac{f^{+}_{i p}
f_{i p'}}{ (1 + \delta R_{i})} \frac{f^{+}_{j p'} f_{i p}}{ (1 +
\delta R_{j})} + Nr_{0}^{2}  \sum_{<ij>}  (V_{ij}-
\frac{1}{2}J_{ij})  \delta R_{i}\delta R_{j}  \nonumber\\
 & +
&\sum_{i,p} f^{+}_{i p}f_{i p}\frac{1}{(1 +  \delta R_{i})}\;
\delta{\lambda_{i}}+ L_{ghost}.
\end{eqnarray}
\end{widetext}

$(sdet M_{AB})^{\frac{1}{2}}$ leads to 
$L_{ghost}({\bf {\cal Z}}) = - \sum_{i p}\; {\bf {\cal
Z}}^{\dag}_{i p}\left(\frac{1}{1 + \delta R_i}\right) {\bf {\cal
Z}}_{i p}$
when is written in terms of complex boson ghost field ${\bf {\cal Z}}_p$
\cite{Foussats}.

Now, we treat the exchange terms $J_{ij}$. These
can be decoupled in terms of the 
bond variable $\Delta_{ij}$ through a Hubbard-Stratonovich transformation,
where  $\Delta_{ij}$ is
the field associated with the quantity
$\sum_{p} \frac{f^{+}_{j p} f_{i p}}{ \sqrt{(1 + \delta R_{i})
(1 + \delta R_{j})}}$.

Finally,  the Lagrangian (4) results

\begin{widetext}
   \begin{eqnarray}
L_{eff}& = & - \frac{1}{2}\sum_{i,p}\left(\dot{f_{i p}}f^{+}_{i p}
+ \dot{f^{+}_{i p}}f_{i p}\right) \frac{1}{(1 + \delta R_{i})} +
\sum_{<ij>,p}\;(t_{ij}\ r_{o}f^{+}_{i p}f_{j p}+h.c.) 
- (\mu - \lambda_{0})\;\sum_{i,p}\;f^{+}_{i p}f_{i p}
\frac{1}{(1 + \delta R_{i})}\nonumber \\ 
&+&
N\;r_{0}\;\sum_{i}\delta{\lambda_{i}}\;\delta R_{i} 
+  \frac{2 N}{J}
\sum_{<ij>}\Delta^{+}_{ij}\;\Delta_{ij}- \sum_{<ij>}\left(\sum_{p}
\frac{f^{+}_{i p} f_{j p}}{ \sqrt{(1 +
 \delta R_{i}) (1 + \delta R_{j})}}\; \Delta_{ij} + h.c.\right)\nonumber\\
& + &  Nr_{0}^{2}   \sum_{<ij>}  (V_{ij}-  \frac{1}{2}J_{ij})
\delta R_{i}\delta R_{j}  + \sum_{i,p} f^{+}_{i p}f_{i
p}\frac{1}{(1 + \delta R_{i})}\; \delta{\lambda_{i}}
+ L_{ghost}.
\end{eqnarray}
\end{widetext}

At this point it is necessary to discuss similarities and differences between 
the SBA and our approach.

a) From step f), we see that the fermions 
$f_p$ are proportional to the 
constrained $X^{0 p}$ operators. 
They are not associated with the spinons 
as in the SBA.

b) From step e), the field $\delta R$ is proportional to the 
real $X^{00}$ operator representing the number of holes (empty sites). 
This is not associated with the holons 
as in the SBA. 

c) The bond variable $\Delta_{ij}$ looks close to the 
valence bond variable of the SBA.
However, $\Delta_{ij}$ besides to be a function of the correlated 
fermions $f_p$, it is also 
a function of $\delta R$ through the denominator. 

Note that $L_{eff}$ (Eq.(5)) contains several nonpolynomial terms.
These apparent complications are the price we have to pay 
for working in terms of 
$X$-operators.
Since the $t-J-V$ model Hamiltonian is quadratic in the $X$'s, 
the strong 
electronic interactions are contained in the commutation rules and the 
constraints.
In the path integral formulation, the information 
contained in the Hubbard 
algebra was transfered to the effective theory.  

In the next sections we will show that present 
theory can go beyond a
formal level and, the obtained results can be compared with others 
in the literature.

Now, we write the  $\Delta_{ij}$ fields in term of static mean-field 
values and dynamics fluctuations 
 $ \Delta^{\eta}_{i}=\Delta (1+r^{\eta}_{i}+i A ^{\eta}_{i})$, where
 ${\eta}=x,y$ and   $r^{\eta}_{i}$ and $A ^{\eta}_{i} $ correspond to the
 amplitude and the phase fluctuations of the bond variable respectively.

To implement the $1/N$ expansion, 
the nonpolynomial $L_{eff}$
should be developed, as in Ref.[\onlinecite{Foussats}],  
in powers of $\delta R$. Up to order $1/N$ 
the following 
Lagrangian is sufficient

\begin{widetext}
\begin{eqnarray}
&L_{eff}&=-\frac{1}{2}\sum_{i,p}\left(\dot{f_{i p}}f^{+}_{i p}
+ \dot{f^{+}_{i p}}f_{i p}\right) (1 - \delta R_{i} + \delta
R_{i}^{2})  
+\sum_{<ij>,p}\;(t_{ij} r_{o} f^{+}_{i
p}f_{j p}+h.c.) - \mu \;\sum_{i,p}\;f^{+}_{i p}f_{i p} (1 - \delta R_{i}
+ \delta R_{i}^{2}) \nonumber \\ & + &
N\;r_{0}\;\sum_{i}\delta{\lambda_{i}}\;\delta R_{i} +\sum_{i,p}
f^{+}_{i p}f_{i p}(1 - \delta R_{i}) \;
\delta{\lambda_{i}} + \frac{2 N}{J}
\Delta^{2}\sum_{i\eta}\left[({r_{i}^{\eta}})^{2}+({A_{i}^{\eta}})^{2}\right]
\nonumber
\\
& -&\Delta \sum_{<ij>,p, p'} (f^{+}_{i p}f_{j p'}+f^{+}_{j p'}f_{i
p})[1 -\frac{1}{2} (\delta R_{i} + \delta
R_{j})+\frac{1}{4}\delta R_{i}\delta R_{j}+\frac{3}{8}(\delta
R_{i}^{2} + \delta R_{j}^{2})] \nonumber
\\ & -&\Delta \sum_{<ij>,p, p'} (f^{+}_{i p}f_{j p'}+f^{+}_{j
p'}f_{i p})({r_{i}^{\eta}}+i {A_{i}^{\eta}})[1 -\frac{1}{2}
(\delta R_{i} + \delta R_{j})] \nonumber
\\
& + &  Nr_{0}^{2}   \sum_{<ij>}  (V_{ij}-  \frac{1}{2}J_{ij})
\delta R_{i}\delta R_{j}
 -\sum_{i p}\; {\bf {\cal Z}}_{i
p}^{\dag}\left(1- \delta R_i+ \delta R^2_i\right) {\bf {\cal
Z}}_{i p},
\end{eqnarray}
\end{widetext}

\noindent
where we have changed $\mu$ to $\mu-\lambda_0$ and dropped
constant and linear terms in the fields.

Looking at the effective Lagrangian (6), the Feynman rules can be
obtained as usual. The bilinear parts give rise to the propagators
and the remaining pieces are represented by vertices. Besides, we
assume the equation (6) written in the momentum space once the
Fourier transformation was performed.

In leading order of $1/N$, we    
associate with the N-component fermion field $f_{p}$, 
connecting
two generic components $p$ and $p'$, 
the
propagator

\begin{eqnarray}
G_{(0)pp'}(k, \nu_{n}) = - \frac{\delta_{pp'}}{i\nu_{n} -
(E_{k} - \mu )}
\end{eqnarray}

\noindent
which is $O(1)$. 

In (7), 
$E_{k}=  
-2(tr_{0}+\Delta) (cosk_x+cosk_y)$, 
is the electronic dispersion in  leading order, where  
$t$ is the
hopping between nearest neighbors sites on the square lattice.

\begin{figure}
\begin{center}
\setlength{\unitlength}{1cm}
\includegraphics[width=7cm,angle=0]{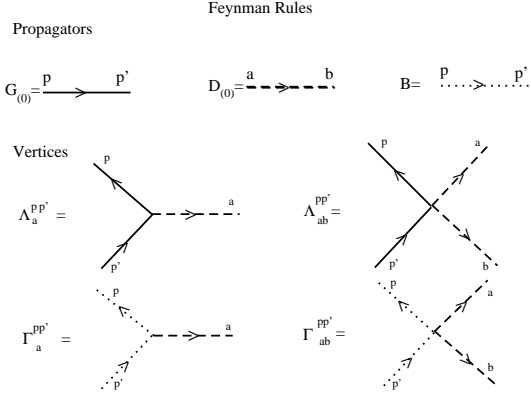}
\end{center}
\caption{Summary of the Feynman rules. Solid line represents the 
propagator $G_{(0)}$ (Eq.(7)) for the correlated fermion $f_p$. Dashed line 
represents the $6 \times 6$ boson propagator 
$D_{(0)}$ (Eq.(8))
for the $6$-component field $\delta X^a$. Note that the component 
$(1,1)$ of this propagator is directly associated 
with the $X^{00}$ charge operator. Doted line is the propagator 
$B$ for the boson ghost field ${\cal Z}_p$. 
$\Lambda^{pp'}_a$ (Eq.(9)) and $\Lambda^{pp'}_{ab}$ represent 
the interaction between two fermions $f_p$ and one and two bosons 
$\delta X^a$ respectively.
$\Gamma^{pp'}_a$ and $\Gamma^{pp'}_{ab}$ represent 
the interaction between two ghost fields ${\cal Z}_p$ and one and two bosons 
$\delta X^a$ respectively.
}
\label{}
\end{figure}

The mean field values $r_0$ and $\Delta$ must be determined 
minimizing the leading order theory. 
From the completeness condition  $r_0$ is
equal to $\delta/2$ where $\delta$ is the hole doping away from
half-filling. On the other hand, minimizing respect to $\Delta$ 
we obtain $\Delta=\frac{J}{2} \frac{1}{N_s} 
\sum_k cos(k_x) n_F(E_k-\mu)$ where $n_F$ is the Fermi function and   
$N_s$ is the number of sites in the Brillouin zone (BZ).

For a given doping $\delta$,  
the chemical potential $\mu$ and $\Delta$  
must be determined self-consistently from 
$(1-\delta)=\frac{2}{N_s} \sum_{k} n_F(E_k-\mu)$. 

We associate with the six component $\delta X^{a}
= (\delta R\;,\;\delta{\lambda},\; r^{x},\;r^{y},\; A^{x},\;
A^{y})$ boson field, the inverse of the propagator, 
connecting two generic components a and b,

\begin{widetext}
\begin{eqnarray}
D^{-1}_{(0) ab}(q,\omega_{n})= N \left(
 \begin{array}{cccccc}
(4V-2J)r_{0}^{2}(cos(q_{x})+cos(q_{y})) 
& r_{0} & 0 & 0 & 0 & 0 \\
   r_{0} & 0 & 0 & 0 & 0 & 0 \\
   0 & 0 & \frac{4}{J}\Delta^{2} & 0 & 0 & 0 \\
   0 & 0 & 0 & \frac{4}{J}\Delta^{2} & 0 & 0 \\
   0 & 0 & 0 & 0 & \frac{4}{J}\Delta^{2} & 0 \\
   0 & 0 & 0 & 0 & 0 & \frac{4}{J}\Delta^{2} \
 \end{array}
\right)
\end{eqnarray}
\end{widetext}

\noindent 

The bare boson propagator $D_{(0)ab}$ 
(the inverse of Eq.(8)) is $O(1/N)$. 
As we will see  
$D_0$ is renormalized to $D$ by 
an infinite series of diagrams of $O(1/N)$. 

We associate with the N-component ghost field ${\bf {\cal
Z}}_{p}$, the propagator,
connecting two generic components $p$ and $p'$, 
$B_{pp'} = - \delta_{pp'}$
which is $O(1)$.

\begin{widetext}

The expressions for three-leg and
four-leg vertices are:

a)

\begin{eqnarray}
\Lambda^{pp'}_{a}& =&  (-1)\; \left(\frac{i}{2}(\nu_n + {\nu'}_n) +
\mu + 2\Delta \sum_{\eta} cos(k_\eta-\frac{q_\eta}{2})\; 
cos\frac{q_\eta}{2};\;1;
- 2\;\Delta\; cos(k_x-\frac{q_x}{2}) \; ; \right.\nonumber \\
&-& \left. 2\;\Delta\; 
cos(k_y-\frac{q_y}{2}) ;\right.
\left. 2\; \Delta\; sin(k_x-\frac{q_x}{2}) \;;2 \Delta\; 
sin(k_y-\frac{q_y}{2}) \right)\;\delta^{pp'}
\end{eqnarray}

\noindent
represents the interaction between two fermions and
one boson.

b)
$\Lambda^{pp'}_{ab}$, 
which represents  the interaction between two fermions and
two bosons, is a $6 \times 6$ matrix where the only elements different
from zero are:

\begin{eqnarray}
\Lambda^{pp'}_{\delta R \delta R}& =&  \left(\frac{i}{2} (\nu_n + {\nu'}_n) 
+ \mu  
+ \Delta\; \sum_{\eta} 
cos(k_\eta-\frac{q_\eta+q'_\eta}{2})\;
[ cos\frac{q_\eta}{2} \; cos\frac{q'_\eta}{2}\;
+\;cos\frac{q_\eta+q'_\eta}{2}] \right)
\delta^{pp'}
\end{eqnarray}
\end{widetext}

\begin{eqnarray}
\Lambda^{pp'}_{\delta R \delta\lambda}=\frac{1}{2}
\;\delta^{pp'}
\end{eqnarray}

\begin{eqnarray}
\Lambda^{pp'}_{\delta R \; r^{\eta}}= -\Delta \; 
cos(k_\eta-\frac{q_\eta+q'_\eta}{2})\; cos\frac{q'_\eta}{2}
\;\delta^{pp'}
\end{eqnarray}

\begin{eqnarray}
\Lambda^{pp'}_{\delta R \; A ^{\eta}}= \Delta \; 
sin( k_\eta-\frac{q_\eta+q'_\eta}{2})\; cos\frac{q'_\eta}{2}
\;\delta^{pp'}
\end{eqnarray}

c)
$\Gamma^{pp'}_{a} = (-1) (\delta_{pp'}\;,\;0,\;0,\;0,\;0,\;0)$
represents the interaction between two ghosts and one
boson. 

d) ${\Gamma}^{pp'}_{ab}$ is a $6\times6$ matrix, 
where ${\Gamma}^{pp'}_{{\delta} R {\delta} R}=\delta_{pp'}$ and, 
the others components are zero. 
It represents the interaction between two bosons and two
ghosts.

Each vertex conserves the momentum and energy and they are $O(1)$.
In addition, in each diagram there is a minus sign for each fermion loop and 
a topological factor. 

Fig.1 summarizes the Feynman rules. 

After identifying the propagators and vertices and the respective
order of them, in the next sections, we will  
calculate different physical quantities. 

Before finishing this section, one remark is necessary. From the $N$-extended 
completeness condition we can see that the charge operator 
$X^{00}$ is $O(N)$, while the operators $X^{pp}$ are $O(1)$. 
This fact will have the physical consequence that the $1/N$ approach 
weakens the effective spin interactions compared to the one related to the 
charge degrees of freedom. This is discussed in 
the next section. 
 
\section{Charge Correlations}

In this section density-density correlations functions are calculated. 
It will be showed that the developed  
formalism can be used in explicit  
calculation of different correlation functions and 
we will also compare our results with others in the literature. 

The density-density correlation function is defined as \cite{Lew,Foussats}

\begin{eqnarray}
\tilde{D}_{ij}=\frac{1}{N} \sum_{pp'} 
<T_{\tau} X^{pp}_i X^{p'p'}_j>
\end{eqnarray}

Using $\sum_p X^{pp}_i=N/2- X^{00}_i$ we find for $\tilde{D}$
in  
the Fourier space 

\begin{eqnarray}
\tilde{D}({\bf q},\omega_n)=
-N {(\frac{\delta}{2})}^2 D_{\delta R \delta R}({\bf q},\omega_n)
\end{eqnarray}

The charge correlation, in $O(1)$, needs the calculation of 
all $O(1/N)$  contributions to 
$D_{\delta R\delta R}({\bf q},\omega_n)$. 
From the Dyson equation, $(D_{ab})^{-1} = (D_{(0) ab})^{-1} - \Pi_{ab}$, 
the dressed  components $D_{ab}$ of 
the boson propagator can be found after 
the evaluation of the $6 \times 6$ boson self-energy matrix $\Pi_{ab}$. 
Using the Feynman rules we may evaluate  
$\Pi_{ab}$  through the diagramas of Fig.2. 
Note that $D_{\delta R \delta R}$ is the element $(1,1)$ of 
the $6\times6$ 
dressed propagator.

\begin{figure}
\begin{center}
\setlength{\unitlength}{1cm}
\includegraphics[width=7cm,angle=0]{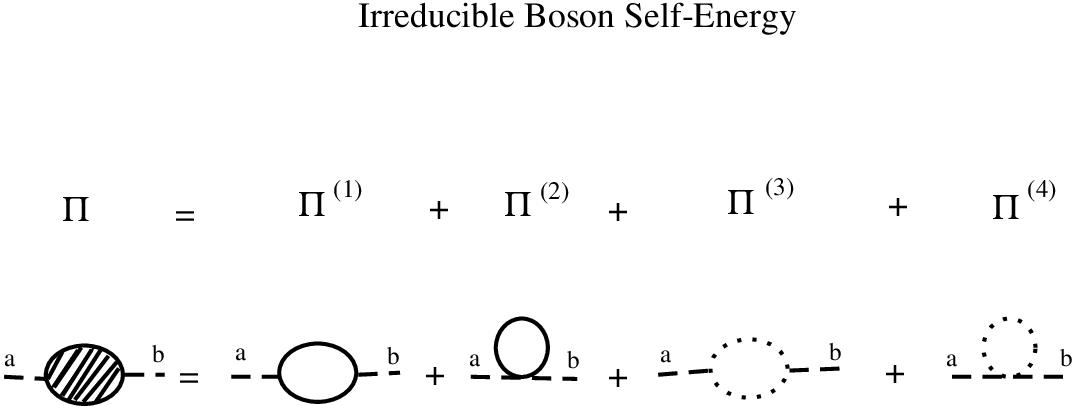}
\end{center}
\caption{The four different contributions 
$\Pi^{(i)}_{ab}$ ($i=1,2,3,4$) to 
the irreducible boson self-energy $\Pi_{ab}$.
}
\label{}
\end{figure}

We note here an important difference with respect to the SBA.
In SBA, only in leading order, the charge-charge 
correlation can be associated with the holon propagator. 
Beyond leading order, the convolution of 
holon propagators (which means the 
reconstruction of the $X$-operator $X^{00}$)
is necessary.  
Meanwhile, in our case, it is not used 
any decoupling scheme and then, the different 
correlation functions are 
directly associated with our field variables.

The density-density spectral function, 
which in principle  can be measure with 
electron energy loss scattering \cite{EELS}, 
is defined as 
$D({\bf q},\omega)=-Im[\tilde{D}({\bf q},\omega)]$. 
The imaginary part is taken as usual after performing the 
analytical 
continuation $i\omega_n=\omega+i\eta$.

In Fig.3 we show the density-density correlation function for the 
physical 
 value $J=0.3$ and for doping $\delta=0.20$. 
We plotted the densities for different ${\bf q}$ vectors in the 
BZ as a function of $\omega$. 
The ${\bf q}$ vectors and the physical parameters are  
the same to those used in the calculation of the densities 
in Ref.[\onlinecite{Zeyher2}]. We used $\eta=0.1$ in the 
analytical continuation.

Comparing Fig.3 with Fig.3 of  Ref.[\onlinecite{Zeyher2}], 
we find a remarkable agreement between both methods in 
spite of the fact that 
the two 
approaches are very different.

\begin{figure}
\begin{center}

\setlength{\unitlength}{1cm}
\includegraphics[width=8cm,angle=0]{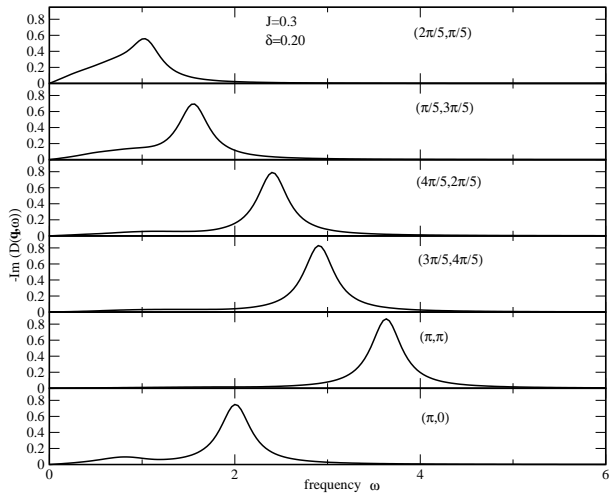}
\end{center}
\caption{Density-density correlation functions for $J=0.3$, $V=0.0$ and 
$\delta=0.20$ for different ${\bf q}$'s in the BZ. 
The density-density correlation functions contain collective peaks 
due to infinite 
diagram summation seen in Fig.2. The collective peaks 
are well pronounced when they are above 
the particle-hole continuum (see for example ${\bf q}=(\pi,0)$). 
They are less pronounced when they are superimposed 
to the particle-hole continuum (see for example ${\bf q}=(2\pi/5,\pi/5)$).
$\omega$ is in units of $2t$ and $t$ is considered to be $1$. 
}
\label{}
\end{figure}

The density-density 
correlation function is  nearly independent of $J$. 
For example, comparing Fig.3 (for J=0.3) with Fig.2 in 
Ref.[\onlinecite{Foussats}](for $J=0$) for momentum 
${\bf q}=(\pi,\pi)$, we find a well 
pronounced collective peak (or zero sound) 
at $\omega \sim 3.5$. 
(Note that in Ref.[\onlinecite{Foussats}] 
$\omega$ is in units of $t$ meanwhile in Fig. 3 it is in units 
of $2t$ in order to absorb the factor $1/N$ due 
to the re-scaling of the 
hopping term.)  

The fact that the density correlations are nearly independent of $J$ 
means that $\tilde{D}({\bf q},i \omega_n)$ is 
dominated by charge fluctuations. 
As we mentioned above, the present formalism  privileges charge over 
spin fluctuations. 
Another consequence of this result is the absence, in $O(1)$,  
of collective excitations (like magnons) in the spin 
susceptibility. 
The spin-spin correlation function is 
the electronic bubble with renormalized band due to 
correlations \cite{Foussats,Lew}. 
Meanwhile there are collective effects in the charge sector in $O(1)$, 
that appear 
in $O(1/N)$
in the spin sector. 

The collective peaks are more pronounced when
they are present  
well above the particle-hole continuum as for ${\bf q}=(\pi,\pi)$. 
For momentum 
${\bf q}=(2 \pi/5, \pi/5)$ the collective peak is superimposed to 
the particle-hole
continuum and it appears broader.
The broadening of the collective peak ${\bf q}=(\pi,\pi)$
is not intrinsic and it is 
only due to the finite value of $\eta$ we used in the analytical continuation.

At this point we can compare  
our results with those obtained using exact 
diagonalization \cite{Tohyama}.
Despite the charge correlation function in Ref.[\onlinecite{Tohyama}] 
was for doping $\delta=0.25$ 
and $J=0.4$ 
and our calculation is for $\delta=0.20$ and $J=0.3$, 
there are some similarities and differences to remark.
For example, Fig. 2 in Ref.[\onlinecite{Tohyama}] 
shows also a peak at ${\bf q}=(\pi,\pi)$.
However,  a) this  peak  
appears at larger energies than our collective peak, 
b) the peak at $(\pi,\pi)$ obtained 
in Ref.[\onlinecite{Tohyama}] contains 
more structure than the ours and  c) we must note also a difference at 
${\bf q}=(\pi,0)$. 
In our calculation we found a collective-like peak at 
$\omega \sim 2$ and 
a broad and small continuum
at low energy. 
Lanczos diagonalization presents also this picture but, 
in addition there is a peak at lower energy. 
As it was already pointed out by Khaliullin and Horsch  
in Ref.[\onlinecite{Kha1}] (see also Ref.[\onlinecite{Kha2}]), 
the inclusion  
of fluctuations beyond the mean field level is probably responsible 
for the differences listed in a), b) and, c).
We think that the inclusion of fluctuations is more important at lower 
than at larger doping. 
In Ref.[\onlinecite{Merino}], for doping 
$\delta=0.5$, we found a better agreement between
the peak position for ${\bf q}=(\pi,\pi)$ obtained by our approach 
and that obtained by Lanczos.

\begin{figure}
\begin{center}
\setlength{\unitlength}{1cm}
\includegraphics[width=7cm,angle=-90.]{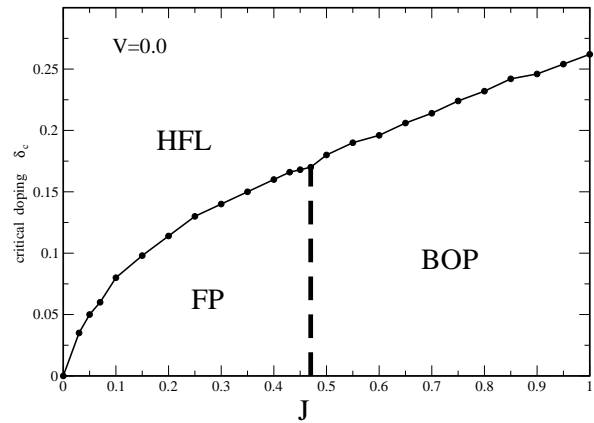}
\end{center}
\caption{Phase diagram in the $\delta_c-J$ plane for $V=0.0$. 
The solid line defines the stability border for the homogeneous 
Fermi liquid (HFL). For a given $J$ the HFL is stable for $\delta>\delta_c$.
The thick dashed line around $J \sim 0.5$ separates the flux 
phase (FP) from the bond order phase  
(BOP). 
}
\label{}
\end{figure}

\section{Instabilities}

\subsection{Flux and Bond-Order phases}

The theory developed in previous sections defines a 
homogeneous Fermi liquid 
(HFL)  
phase.
In $O(1)$ the mean field solution ($r_0$, $\Delta$) was independent of 
sites. Therefore, at $N$-infinite, 
we have free fermions with renormalized 
band $E_k$ due to correlations. 

In this section we will study the stability conditions for the HFL. 

In leading order, there are collective effects in the charge 
sector whereas there are not in the spin sector. 
Therefore, in principle,  we expect 
instabilities in the charge channel. 

The HFL system is unstable when 
the static charge susceptibility 
$Re(\tilde{D}({\bf q},i\omega_n=0)$ 
diverges. The ${\bf q}$ vector, 
where the instability occurs ($\bf{q_c}$), being the modulation 
of the new phase.

Fig.4 shows the phase diagram of the model for $V=0$. 
For a given $J$, below a critical doping $\delta_c$
where the static charge 
susceptibility diverges at a given vector ${\bf q}$ in the BZ,
the HFL is not the stable phase.

In the limit $J \rightarrow 0$ 
$\delta_c \rightarrow 0$ the HFL is stable for 
the whole doping range except 
at half filling
where the system is an insulator because the band effective-mass  
tends to infinite.

The instability is placed, for all $J$,  on  the border of the BZ. 
That is,  ${\bf q_c}=(1,x) \pi$
or ${\bf q_c}=(x,1)\pi$ 
where  
the parameter $x$ measures the degree of the  
incommensuration of the instability. 

In Fig.5 we plot the incommensuration $x$ as a function of $\delta_c$.
For $\delta_c \rightarrow 0$, $x \rightarrow 1$. Therefore, in 
the limit of $J \rightarrow 0 $,  we found a phase 
with commensurate 
order ${\bf q_c}=(\pi,\pi)$.

It is easy to overlook the instability just by looking at the 
static charge susceptibility. 
In Fig.6 we plot the static charge susceptibility, along 
${\bf q}=(\pi,q)$, for 
$\delta=0.15$ and $\delta=0.14$ and,  for the physical value $J=0.3$.
From Fig.4 we know that the HFL is not stable for doping $\delta=0.14$
whereas  for $\delta=0.15$ it is stable.  
Both curves in Fig.6 look similar and, for $\delta=0.14$ 
there is no indication of the instability for  ${\bf q} \sim {\bf q_c}$ 
because it occurs 
in a very narrow region of momentum near ${\bf q_c}$. 
In practice,   
this means that the
set of ${\bf q}$-points in Fig.6 is 
not dense enough to localize the divergence of 
$Re(\tilde{D}({\bf q},i\omega_n=0))$.  
As we will see below, this is related with the fact that the instability 
is weakly coupled with the charge sector. 

\begin{figure}
\begin{center}
\setlength{\unitlength}{1cm}
\includegraphics[width=8cm,angle=0]{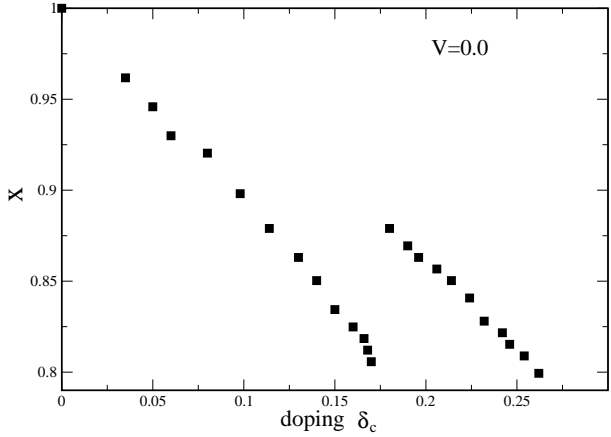}
\end{center}
\caption{Inconmensuration $x$ vs $\delta_c$ for $V=0.0$.
$x$ is defined as ${\bf q_c}=(1,x)\pi$. 
${\bf q_c}$ is the ${\bf q}$-vector 
where the instability takes place.
The instability is incommensurate for all $J$ except 
for $J \rightarrow 0$
where $x \rightarrow 0$ (${\bf q_c} \rightarrow 
(\pi,\pi)$). At the separating border of the FP from the BOP there is a 
jump in the value of ${\bf q_c}$. 
}
\label{}
\end{figure}

To calculate $\tilde{D}({\bf q},\omega_n)$ 
we have to evaluate the inverse 
of $(D_{ab})^{-1}$.  
Therefore, for a better determination of the instability,  
we may look for the zeros of the 
determinant of $D_{ab}^{-1}$.
In the inset of Fig.6 we plot the $det(D^{-1}_{ab})$ as a function of 
${\bf q}=(\pi,q)$. For $\delta=0.14$ 
(close to the onset of the instability)  
the determinant changes 
sign in a narrow region around ${\bf q_c} \sim (\pi,2.7)$ making the HFL 
unstable. Meanwhile, for 
$\delta=0.15$ the determinant is positive for all {\bf q} 
and the system is stable.
A similar plot for $\delta < 0.14$ shows, 
of course, a larger negative region
for $det(D_{ab}^{-1})$.

In order to characterize the nature of the new phase, we have studied 
the eigenvector corresponding to the zero eigenvalue of 
$(D_{ab})^{-1}$. 

For doping $\delta < \delta_c \sim 0.5$ 
the eigenvector corresponding to the zero mode 
is $\sim (0,0,0,0,-1,1)$. Then, in the new phase,  
the 5th and 6th components 
of $\delta X^a$ ($A^x$ and $A^y$) are frozen. 
$A^x$ and $A^y$ are associated with the phase 
of the field $\Delta^{\eta}$
and they take opposite values. 
Therefore,  as the instability occurs near the momentum $(\pi,\pi)$, 
this new phase is the well known flux phase (FP) 
\cite{Marston,Capellutti}
which opens a gap, 
with $d$-wave symmetry
in the normal state.

Recently, 
Chakravarty, Laughlin, Morr and Nayak 
considered that the FP
or DDW as a candidate to explain the physics 
of the pseudogap 
in underdoped cuprates\cite{Chakra}. 

The first component of the eigenvector corresponding 
to zero eigenvalue 
is not zero exactly. It has a small value which 
means that the FP is weakly
coupled to the charge sector. 
It means that the DDW is not completely hidden when it is 
incommensurate. Although Bragg peaks 
are predicted, they will show low intensity which would make their 
observation very difficult. 
This is the reason by which,  
any charge-probe is not very sensitive 
to show the FP instability.
When ${\bf q_c}$ is exactly $(\pi,\pi)$ 
the eigenvector does not have any  
mixing with the charge sector and, the flux phase is fully hidden.

For doping $\delta > \delta_c \sim 0.5$ the unstable 
eigenvector is of the form 
$\sim (0,0,0,1,0,0)$. This state is known  as  bond-order phase (BOP)
\cite{Capellutti,
Lubenski}.

In the range of the  studied parameters we did not find indications 
for phase separation.

In Fig.5,  
for $J \sim 0.5$, near the crossover from the 
FP to BOP, there is a  jump in the 
incommensuration $x$.

The agreement with other methods 
\cite {Capellutti,Lubenski} 
is again remarkable in spite of the fact that the
present approach is, {\it a priori},  very different from those ones.
Taking into account that we constantly find 
similar or even the same results 
by means of different methods turn the results reliable.  
For example,  
for the physical value $J=0.3$, the HFL is unstable against 
a FP. 
This result is very robust and,  many different 
methods agree that the stable 
phase, for small doping, is a FP for the physical 
region of the parameters
\cite{Marston,Grilli,Lubenski,Capellutti,
Numerics}.

\begin{figure}
\begin{center}
\setlength{\unitlength}{1cm}
\includegraphics[width=8cm,angle=0.]{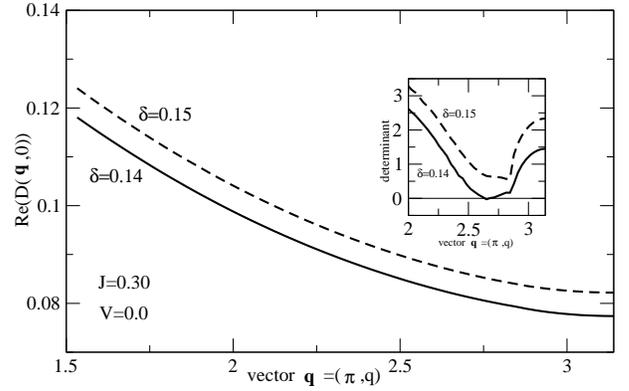}
\end{center}
\caption{Static charge susceptibility 
$Re(\tilde{D}({\bf q},i\omega_n=0))$
vs ${\bf q}=(\pi,q)$ for 
$J=0.3$, $V=0.0$ and, for $\delta=0.15$ and $\delta=0.14$.
For $\delta=0.14$ (solid line) the curve looks similar 
to the case for $\delta=0.15$ (dashed line). For $\delta=0.14$ the HFL 
is not stable and there is 
no strong sign  of the instability in the static charge susceptibility.
In the inset we plot the $det (D_{ab}^{-1})$ (see text for discussions) 
vs ${\bf q}=(\pi,q)$. For $\delta=0.14$, the $det (D_{ab}^{-1})$ clearly 
changes sign 
in a small region around ${\bf q_c} \sim (\pi,2.7)$.
}
\label{}
\end{figure}

\subsection{Charge Density Wave phase} 

The inclusion of a nearest-neighbors Coulomb repulsion $V$ 
favors a charge density wave (CDW) state. 
In order to investigate this instability,
we have included a finite value of $V$ in the calculation.

The study of a nearest-neighbors Coulomb repulsion 
in correlated models is important.
In cuprates,  there are indications \cite{Fein} showing the existence 
of $V$ with a value close to $J$. 
If $J$ favors superconductivity and,  the value of $V$ is large, 
it is reasonable to think 
that superconductivity will be diminished due the 
presence of $V$.
There are some analytical and numerical works 
studying the competition between $V$ and $J$ on superconductivity. 
Meanwhile some papers indicate that superconductivity in the $t-J$ model 
vanishes for $V \sim J$ \cite{Zeyher1}, others indicate that 
superconductivity survives 
up to values of $V >> J$ \cite{Riera}.

On the other hand, the presence of $V$ seems to be important 
for understanding the physics of organic materials \cite{jaimes}.
Some peculiarities in the optical conductivity 
(and also superconductivity) 
occur near the charge order and this  
picture can be interpreted by
the competition between $V$ and the kinetic energy in correlated 
models \cite{org}. 

Looking at the Eq.(8) we see that $V$ is only present  
in the element $D^{-1}_{(0)}(1,1)$. 
The Coulomb term enters in our approach
multiplied by $(\delta/2)^2$  which means that the effect of $V$, 
at low doping, 
is strongly 
screened by 
the correlations.  

We note that in the SBA $V$ can enter in different manners  
depending on the way chosen for the decoupling.

In Fig.7 we show the phase diagram in the $V_c-\delta$ plane 
for J=0 and $J=0.3$. 
The curves 
show,  as a function of $\delta$,  the 
critical Coulomb repulsion $V_c$ where the CDW instability takes place.  
For $V <V_c$ the HFL is stable. 
This result is in agreement with numerical \cite{Merino} and 
analytical methods 
\cite{Hoang}. 
In Ref.[\onlinecite{Hoang}] the authors use coherent 
potential approximation and 
their results are 
close to the ours. 

A similar result was also recently found, at $J=0$, in the 
triangular lattice in the  context of the new 
low dimensional superconductor 
$Na_xCoO_2$ \cite{Lee}.  

\begin{figure}
\begin{center}
\setlength{\unitlength}{1cm}
\includegraphics[width=7cm,angle=-90.0]{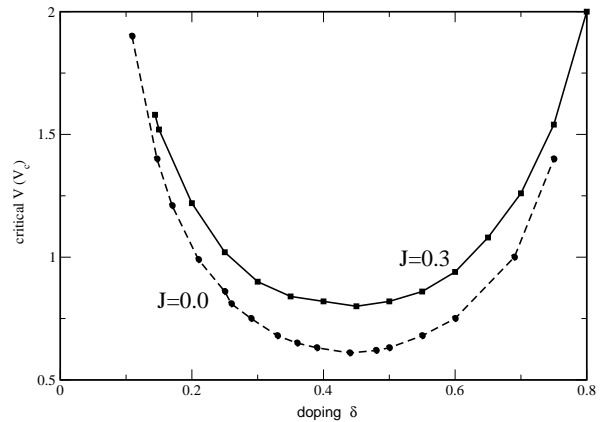}
\end{center}
\caption{Charge density wave (CDW) stability line $V_c$ vs $\delta$
for $J=0.0$ (dashed line) and $J=0.3$ (solid line). For a given $\delta$,
the system is in a CDW state for $V > V_c$. The value of $V_c$ increases 
with increasing $J$.
For $J=0.3$ the first presented doping is $\delta=0.15$. 
(For $J=0.3$ the HFL is not stable for $\delta <0.14$
at $V=0.0$). 
}
\label{}
\end{figure}

For $J=0.3$ and doping $\delta=0.20$, 
the HFL is stable at $V=0$. 
When $V= V_c\sim 1.25$
the system enters in a CDW state. This  
$V_c$ is $\sim$ $40 \%$ 
larger than for $J=0$
and the same doping. 
There are two sources
for 
this tendency.
 
a) When $J$ is finite, there is a contribution $\Delta$ to the effective 
hopping in $E_k$. Then, the system win kinetic energy when $J$ 
is finite and,  a larger $V$ is necessary 
in order to localize the charges.

b) As we see in the element $(1,1)$ of $D^{-1}_{(0)}$, the effect of $V$
is diminished 
when $J$ is finite. 
The term $2J$ in the element $(1,1)$ comes 
from the charge-like term 
$J_{ij}X_{i}^{\sigma
\sigma} X_{j}^{\bar{\sigma} \bar{\sigma}}$
of the pure $t-J$ model, which is 
of the same form of the Coulomb term (see Eq.3).

In order to identify the main source for the increasing of $V_c$ 
we looked for
the CDW instability without the charge-like term in (3). 
Under this condition, for $J=0.3$ and $\delta=0.20$, 
$V_c$ is $V_c \sim 1.0$. This value is nearly the 
same than $V_c$ for $J=0$. Therefore,  b) is the main source 
for the increasing of $V_c$ with increasing $J$.     

In contrast to the FP and BOP, 
the CDW instability can be detected clearly 
by looking at the static  charge susceptibility  
$Re(\chi({\bf q},i\omega_n=0))$. 

In Fig.8 we  
show, 
for $J=0.3$ and $\delta=0.25$, 
$Re(\chi({\bf q},i\omega_n=0))$ vs ${\bf q}=(q,q)$ 
for $V=1.$ (a 
little smaller than the critical value $1.1$) and $V=0.5$.    
In the figure, we see that approaching $V_c$, the static 
susceptibility tends to diverge near ${\bf q}=(\pi,\pi)$.
For $V=0.5$ ($V<<V_c$), the static charge susceptibility is flat.

\begin{figure}
\begin{center}
\setlength{\unitlength}{1cm}
\includegraphics[width=7cm,angle=-90.0]{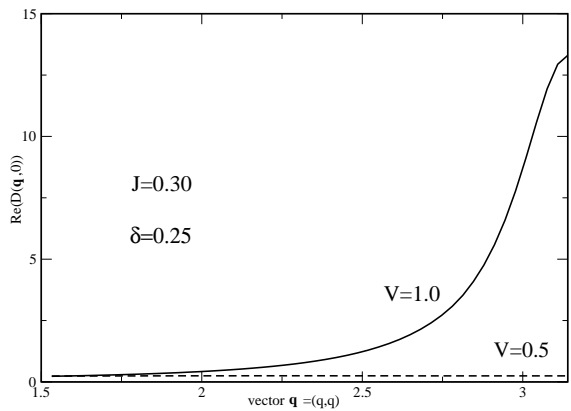}
\end{center}
\caption{Static charge susceptibility for $J=0.3$ and $\delta=0.25$ vs 
${\bf q}=(q,q)$ for two different values of $V$, $V=1.$ and $V=0.5$.
$V=1.$ is near (from below) the value of $V_c=1.1$. 
In contrast to Fig.6, the CDW instability has a clear signal in the 
static charge susceptibility. For $V=1.$ there is a clear 
indication of a divergence at $(\pi,\pi)$.  
For $V=0.5$ ($V<<V_c$), the static charge susceptibility is flat.
}
\label{}
\end{figure}

In this case the eigenvector corresponding to the zero eigenvalue is 
$ \sim (1,0,0,0,0,0)$. 
Then, the instability is mainly in the pure charge sector.
  
From Fig.8, in agreement with other methods \cite{Merino},
the instability  
${\bf q}$-vector is always at ${\bf q}=(\pi,\pi)$. 
Therefore, in the CDW phase,  the charges order themselves 
forming a checkerboard pattern. 
For doping different to the commensurate one $\delta=0.5$ we must 
interpret the CDW instability as the system is charge-ordered 
but remains metallic. 
These results were obtained by different methods and the 
nature of the instability for doping different to the commensurate 
one is still open.
For the commensurate doping $\delta=0.5$  in Ref.[\onlinecite{Merino}] 
we showed that the system is 
charge-ordered at $V_c \sim 0.6$ however, we also find 
an insulator because the quasiparticle 
weight goes to zero exactly at this $V_c$ value.

In order to compare the above results to  
the predictions of the BKF method, we calculated 
the critical $V_c$ using the formulation of Refs.[\onlinecite{Zeyher}] 
[\onlinecite{Zeyher1}] for $J=0.3$. 
For $\delta=0.20$  and $\delta=0.60$ we found 
$V_c\sim 1.25$ and $V_c\sim 0.9$ 
respectively. The agreement with our calculation is again very good
in spite of the fact that 
the Coulomb term enters BKF  
in a different way.   

\section{Conclusions}

From the path integral representation 
for the Hubbard 
$X$-operators we have developed a new perturbative theory for the $t-J-V$ 
model. 
Our formulation is free of slave-particles
and the $X$-operators are treated as fundamental objects.

We have extended the large-$N$ formalism, recently developed in Ref.
[\onlinecite{Foussats}], to the case of finite $J$.

To work directly with the $X$-operators has several advantages upon the 
slave-particles methods. 
For example, in the slave-boson theory, 
the decoupling procedure introduces a 
gauge degree of freedom and then, an additional problem as the gauge 
fixing arises.
This problem does not occur in our formulation.
On the other hand, in the slave approaches, in contrast to our method,
beyond a mean field level, a convolution between 
slave-particles propagators 
is necessary to reconstruct the original 
physical propagator.

Fig.1 summarizes the Feynman rules of the present approach 
where the propagators and vertices are written in terms of the 
Hubbard operators $X^{\sigma 0}$, $X^{00}$, etc. 

It is important to note that  
we can not read the interaction vertices from the  
$t$ and $J$ terms of the Hamiltonian.    
They arise from the nonpolynomial effective theory defined by $L_{eff}$
in Eq.(6). 
All the interactions caused by the 
Hubbard algebra (commutation rules and constraints) were transfered to 
$L_{eff}$. 

The formalism developed in the present paper privileges 
charge over spin fluctuations.
While there are collective effects in the charge sector in $O(1)$, 
those appear in $O(1/N)$ in the spin sector.

We have studied 
charge correlations functions and we have investigated the role of 
collective effects.
This study shows the presence of collective peaks (or zero sound)
which are more evident when they are separated from the 
particle-hole continuum. On the other hand, the collective peaks become 
broader when they are superimposed to the particle-hole continuum.
One important characteristic of the charge correlation function is the 
fact that this is not strongly dependent on the value of the 
exchange interaction $J$.

At large $N$, the theory can be described  as a homogeneous Fermi liquid 
with renormalized band due to correlations.
We have studied the stability of this Fermi liquid phase. 
For a given value of $J$, the Fermi liquid phase is stable for doping 
$\delta > \delta_c$. The critical doping $\delta_c$ decreases with 
decreasing $J$. 
For doping below $\delta_c$ the system enters in a flux or 
bond order phase 
depending if $J < 0.5$ or $ J > 0.5$ respectively. 
These instabilities are 
weakly coupled to the charge sector which means that any charge-probe 
is not an efficient test to detect them. 
One important characteristic of these new phases is that they are 
incommensurate with a modulation vector ${\bf q}=(1,x)\pi$ where the 
incommensuration $x$ tends to $1$ when $J$ goes to zero.

It is important to remark that in agreement with other theories, 
for low doping, the Fermi liquid is unstable 
against a flux or $d$-density wave phase for the physical region of 
the parameters.

We also have investigated the role of a nearest-neighbors 
Coulomb repulsion 
$V$ on the stability conditions of the Fermi liquid.
When $V$ is larger than a critical value $V_c$, the system enters, 
at a given doping, in charge density wave 
state. The value of $V_c$ increases 
with increasing 
$J$. We have identified in the charge-like term of the pure $t-J$ model
the main reason for this increase.  

We have continuously compared our results with similar ones 
in the literature. 
The agreement of our results with those obtained by 
other methods give 
confidence to the results and the approach of the present paper. 

At leading order, our formalism is in agreement with the slave-boson 
approach. However, at the next to leading order 
(which is necessary to calculate dynamical properties) 
the differences between the two 
formulations are not yet completely established. 
We think that the complications that appear in 
the slave-boson method beyond mean field level 
as gauge fixing, Bose condensation, 
regularizing factors \cite{Arrigoni} are not good 
for the advance of the field.
In our case we do not have these problems and we think that our approach 
can be useful to go beyond the mean field level. 
With the formulation for finite $J$ in hand, 
we expect to continue in this direction as we did 
for $J=0$ in Ref.[\onlinecite{Merino}].

\noindent{\bf Acknowledgments}

The authors thank to C. Genz, 
L. Manuel, J. Merino and R. Zeyher 
for valuable discussions.




\begin{thebibliography}{99}

\bibitem{Anderson} P.W. Anderson, The Theory of Superconductivity in
High-$T_c$ Cuprates (Princeton University Press, Princeton,1997).

\bibitem{McKenzie} R.H.McKenzie, Science {\bf 278},820(1997).

\bibitem{Izyumov} A. Izyumov, Physics-Uspekhi {\bf 40}, 445 (1997).


\bibitem{Dagotto} E. Dagotto, Rev.Mod.Phys. {\bf 66},763(1994). 

\bibitem{Hubbard} J. Hubbard, Proc. R. Soc. London Ser. A {\bf 276},
238 (1963). 

\bibitem{Arrigoni} E. Arrigoni et al, Phys. Rep.{\bf 241}, 291(1994). 

\bibitem{Wang} Z. Wang, Int. Journal of Modern Physiscs B {\bf 6}, 155(1992).

\bibitem{Zeyher} R. Zeyher and M. L. Kuli\'c, Phys.Rev. B {\bf 53}, 2850(1996).

\bibitem{Zeyher1} R. Zeyher and A. Greco, Eur. Phys. J.B {\bf 6}, 473 (1998).

\bibitem{Foussats1} A. Foussats, A. Greco, C. Repetto, O. P. Zandron and O. S. 
Zandron, Journal of Physics A {\bf 33}, 5849(2000).

\bibitem{Foussats} A. Foussats and  A. Greco; Phys.Rev.B{\bf 65},
195107(2002).

\bibitem{Merino} J. Merino, A. Greco, R. H. McKenzie, and M. Calandra,
Phys. Rev. B {\bf 68}, 245121(2003). 


\bibitem{Lew}  L. Gehlhoff and R. Zeyher, 
Phys. Rev. B{\bf 52}, 4635(1995).

\bibitem{EELS} E. Zojer, M. Knupfer, Z. Shuai, J. Fink, J.L. Bredas, 
H. H. Horhold, J. Grimme, U. Scherf, T. Benincori, and G. Leising, 
Phys.Rev.B {\bf 61},16561(2000). 


\bibitem{Zeyher2} R. Zeyher and M. L. Kuli\'c, Phys.Rev. B {\bf 54}, 8985(1996).


\bibitem{Tohyama} T. Tohyama, P. Horsch and S. Maekawa, Phys. Rev. Lett.
{\bf 74}, 980(1995).

\bibitem{Kha1} G. Khaliullin and P. Horsch; Phys.Rev. B {\bf 54}, R9600(1996).

\bibitem{Kha2} P. Horsch and G. Khaliullin; cond-mat/0312561 (2003).

\bibitem{Marston} I. Affleck and J.B. Marston, Phys.Rev.B{\bf37},3774(1988).

\bibitem{Capellutti} E. Cappelluti and R. Zeyher, Phys.Rev.B{\bf 59},
6475(1999).

\bibitem{Chakra} S. Chakravarty, R. B. Laughlin, D.K.Morr, and Ch. Nayak, Phys.
Rev. B {\bf 63},94503 (2001). 


\bibitem{Grilli}  M. Grilli and G. Kotliar, Phys.Rev. Lett. {\bf 64}, 1170 
(1990).

\bibitem{Lubenski} D.C. Morse and T.C. Lubensky, 
Phys.Rev. B {\bf 42},7994(1990).


\bibitem{Numerics} P.W.Leung, Phys. Rev. B{\bf 62},6112(2000).  



\bibitem{Fein} L.F. Feiner,J.H.Jefferson and R. Raimondi, Phys.Rev. B {\bf 53}
8751 (1996). 

\bibitem{Riera} J. Riera and E. Dagotto, Phys.Rev.B {\bf 57},8609 (1998);
M. Calandra and S. Sorella, Phys. Rev. B{\bf 61},11894(2000); C. Gazza, 
G.B. Martins, J. Riera and E. Dagotto, Phys.Rev.B {\bf 59}, 709(1999).

\bibitem{jaimes} R.H.McKenzie,J.Merino,J.B.Marston and O.P.Sushkov, 
Phys.Rev.B{\bf 64},085109(2001); J.Merino and R.H.McKenzie, Phys.Rev. Lett.
{\bf 87},237002(2001). 

\bibitem{org} M.Dressel, N.Drichko, J.Schlueter, and J. Merino, 
Phys.Rev.Lett.{\bf 90},167002(2003). 

 
\bibitem{Hoang} A.T. Hoang and P.Thalmeier, J. Phys.: Condens. Matter 
{\bf 14}
6639(2002).

\bibitem{Lee} O. Motrunich and P. Lee, Phys. Rev.B{\bf70},024514(2004). 


\end{thebibliography}
\end{document}